# Intrinsic Awareness,

# The Fundamental State of Consciousness[1]

Weili Luo
Department of Physics, University of Central Florida, Orlando, Florida, 32816, USA
(Weili.luo@ucf.edu)

**Abstract**
In an effort to simplify the complexity in the studies of consciousness, the author suggests to describe the conscious experiences as a fundamental state, the intrinsic awareness (I.A.), and functions of this fundamental state. I.A. does not depend on external environment, our sense organs, and our cognitions. This ***ground state*** of consciousness is timeless and irreducible to sub-constituents; therefore reductionism can apply neither to the analysis nor to the new theory of I.A. The methodology for investigating I.A. is proposed and the relation between I.A. and the hard problem in consciousness proposed by Chalmers is discussed.

## 1. Introduction to Intrinsic Awareness (I.A.)

The question of how the mind or consciousness works has been an important intellectual pursuit of philosophers and psychologists over centuries and even millennia. With the advance of technology, it has gone beyond a philosophical issue and has become one of the most studied topics in science. Because the question encompasses diverse fields such as psychology, philosophy, biology, cognitive science, neuroscience, artificial intelligence, etc., it is a daunting task to even just consider where to start to address the issue of consciousness: should one start from cognition? Should one start from neurobiology, or from the philosophical question of whether the mind is the emergent phenomenon of the physical brain or not? Although different expertise all has something to contribute, the answer to these questions has been entangled in the complexity of many areas overlapping with the problem [1]. In 1996, David Chalmers proposed that there are two types of problems in the studies of consciousness [2], the easy problems, which mainly address the objective mechanisms of the cognitive system, and the hard problem that involves how physical processes in the brain give rise to subjective experiences. The easy problems, although many of them were unsolved then

---

and still are as of today, are solvable; while the hard problem is the one that researchers do not know where to start, and Chalmers was not even sure that current scientific framework can eventually solve it.

In this article a proposal is made, instead of dividing the consciousness studies into the easy problems and hard problem, to characterize our conscious experience by a fundamental state and functions of this state. I argue that the complexity of the studies of consciousness can be simplified because of the existence of a fundamental awareness. I terms it the "Intrinsic Awareness" or I.A., which is our basic ability to be aware, to know. The idea of this fundamental awareness is borrowed from some ancient contemplative traditions where it has been called different names such as innate basis of mind, pure mind, illuminating mind, or simply awareness, etc. [3-6].

It is not difficult to realize that I.A. does not depend on our senses and the external conditions. Taking the example of seeing red roses: the whole process of seeing red roses can be broken down to several stages: we may first sense something beautiful in our environment, then our vision comes into focus so we can see the actual flowers; our brain starts to perform the function of analyzing and comparing the observed object with other concepts stored in it; afterward the brain comes up with the conclusion that this is a bunch of red roses. If these flowers are outside our sight at the beginning, our olfactory organ will be the first to get in contact with the smell from them. And, the analytical process from the brain leads to the conclusion that these are fragrant flowers. Although the cognitive process seems to start from the contact of the object with our visual or olfactory sense, it is our ability to know that leads us to notice that there is something out there in the first place. This ability itself does not depend on external objects and it must exist before our sense organs get in contact with the environment. A blind person may not see things at this moment but his/her ability to see is still there. When technology is advanced enough to cure the blindness the person is no longer blind to outside world, and then the ability to see will be able to perform its functions. Just recently a device has been invented to help blind people to analyze the electric signals received so they can "see" the environment surrounding them[2]. Similarly, people who could not hear may have problems with their sense organ, the ear(s) in this case, but they do not really lose their ability to hear. With modern technology, it is not hard to conclude that once a hearing aid is put into place, they are able to hear again, just as we may not be able to hear in a soundproof room but our ability to hear is still there. Once we get out of the room, we are able to hear again.

The next example goes beyond our sense organs: suppose we are alone in a room with our back facing the door when someone walks into the room quietly. Even we do not hear steps and do not see the person, many of us have the gut feeling that someone else is in the room. Through some training, most people can achieve the same ability[3] associated with I.A. It is also easy to understand that I.A. can function perfectly well without external stimuli — we can be aware of internal activities such as our thoughts, emotions, pains, etc.; we know when we are hungry, thirsty, or when we are in love. These examples demonstrate that our ability to be aware depends neither on the external environment nor on our sense organs.

---

[2] "Acoustics help blind people see the world" The Medical News, July 5, 2009. URL: http://www.news-medical.net/news/20090705/Acoustics-help-blind-people-see-the-world.aspx

[3] One way to achieve this is to train our mind to be tranquil so as to be sensitive to surrounding environment.

## 2. All Conscious Experiences Are Functions of I.A.

If I.A. does not depend on sense organs, the consciousnesses associated with the five sense organs[4] must not be fundamental. Then, what is the relation between I.A. and those conscious experiences we are more familiar with? It is common sense that our daily activities depend on awareness. We have to be aware of our environment and of ourselves to live a normal life, implying that, rather than separating from I.A., our everyday experiences are expressions of I.A. through a specific sense organ or mental activities. When I.A. manifests through our visual organ, the eyes, it becomes what we know as the visual consciousness; when I.A. expresses itself through the hearing organ, it becomes what we know as the hearing consciousness; the same is true for smell, taste, and tactile consciousnesses.

Besides the five sense organs we also have consciousness related to mental activities such as thoughts, concepts, and being conscious of self, etc. Some people, especially those who believe that mind is the emergent phenomenon of brain, may conclude that I.A. must be the product of our mental activities. In the previous example of being aware of someone walking into the room, one learned that I.A. does not depend on the mental discriminative ability based on concepts. At the very first moment, before we make connection with any concept or even any sense organ, we just have a hunch that someone is in the room. In fact, we all have experiences in which we had a hunch about something or some situation that we could not quite put it into words. This realization that we know or feel something but could not express it indicates that 1) our language is not sufficient to describe what we experience; 2) the "inner knowing" comes before any concept or words. This "knowing" before the conceptual mind arises is our basic ability to know, i.e. our intrinsic awareness, I.A. These examples demonstrate that this fundamental awareness is always with us whether we experience something or not. This leads to the conclusion that I.A. is independent of, instead of the product of, mental activities.

Because all our conscious experiences, either through physical senses or mental cognitions such as thinking, perception, judging, remembering, and problem solving, depend on the instinct ability to be aware, none of them is fundamental – they are all functions of I.A.

## 3. Timeless and Non-Reducible I.A. as the Ground State of Consciousness

The fact that I.A. is the foundation to all conscious experiences indicates that it cannot be further divided into sub-constituents: it is neither a part of anything else nor an entity depending on anything else; I.A. is irreducible. Thus, the current reductionist approach, where a system is analyzed by its subunits plus the interactions between them, will not apply to the study of I.A. Inasmuch as I.A. is irreducible, I.A. does not need the concept of "space coordinate", which is based on the potential divisibility, as in the concepts of left and right, up and down, inside and outside, etc.[5] [7]. Because I.A. cannot reduce itself to something else and because I.A. is always with us, nothing changes in this state. Therefore, there is no "time" if one resides in this pure awareness. Apparently, I.A. is the **_ground state_** of consciousness.

---

[4] The five consciousnesses associated with the five sense organs are: the visual consciousness, the auditory consciousness, the nasal consciousness, the taste consciousness, and the tactile consciousness.

## 4. How to Investigate I.A.

If one accepts that I.A. is the root awareness from which all conscious experiences emerge and I.A. is not the product of physical and mental activities, one has to acknowledge the possibility that physical instruments we use in modern science, may not be suitable to measure properties of I.A. *directly*. It is very likely that these apparatus, at the best, probe only the results of interaction between the mental activities in the brain and I.A. Thus the statement "you do not have experimental evidence to show that I.A. is independent of mental activities" is not a valid argument against the existence of I.A.

In quantitative science such as physics, we develop theories through mental construct based on concepts, fundamental laws, and experimental facts; or the other way around, we perform experiments to test theoretical predictions. In all these activities, we use our discriminating ability to differentiate right and wrong, reasonable versus unreasonable. We use concepts, knowledge, even experience stored in our memory to compare with the object under study. There is, however, nothing in our experience or concepts that we learned through analytical or reductionist methodology nearly resembles I.A. Then, the question is how do we learn and understand I.A.? During many years of investigating the nature of consciousness, I realized that, to truly know or even to get familiar with I.A., one must directly experience it through introspection [8] or contemplative practice[6]. In this direct perception our stereotype, culture background, or special training that leads to a special way of thinking, have to be put aside so they will not interfere with the experience.

Although this introspection may seem rather subjective, therefore lack of scientific objectivity, this in fact is not the case. In exploring I.A., as long as we follow the same procedure step by step, different people should have the similarly reportable result about the existence of I.A., albeit the details about how to reach that conclusion may differ. Therefore, the seemingly subjective introspection is verifiable.

---

[5] One should be careful not to equate the space in space-time and the spaciousness. When we say: "someone's mind is as vast as space" we mean spaciousness, not the space coordinate in space-time framework used in modern science.

[6] The contemplative practice is no longer a taboo in our society, even in higher education. Numerous research works have been done to investigate the effect of contemplative practices [9]. Some laboratories and centers for contemplative studies have been established in higher education such as: The Center for Investigating the Healthy Mind at the University of Wisconsin (URL: http://www.investigatinghealthyminds.org); The Association for Contemplative Mind in Higher Education (URL: http://www.acmhe.org); Contemplative Studies Initiatives at Brown University, Emory University, and other institutions.

## 5. I.A. Versus "The Hard Problem" in Consciousness

There are commonalities between "the hard problem" of consciousness proposed by David Chalmers [2] and I.A. discussed by the author of this article. According to Chalmers, the mechanism associated with the interaction between human and information received by the subject, i.e. the cognitive process, and the mechanisms related to it, are the easy parts of consciousness study, meaning at least one can expect to solve them sometime down the road. The subjective experience is hard because one does not even know where to start to address it. In fact one is not even sure that science can provide an answer to the hard problem in consciousness. Chalmers demonstrated the difficulties encountered by reductionist approach when dealing with the hard problem in consciousness. As far as the inapplicability of the reductionism to the underlying problems is concerned, I.A. and the hard problem proposed by Chalmers face the same situation. Nevertheless, characterizing conscious experience as I.A. and functions of I.A. actually simplifies the complexity involved in studies of consciousness. Once we know the fundamental state of the consciousness we know where to start to proceed. Furthermore, this work gives a clear answer to the question "why doesn't the reductionism apply to the 'hard problem' in consciousness study?"

There are clear distinctions between the "hard problem of consciousness" and I.A.: most of David Chalmers' examples as the hard problem of consciousness are subjective experiences that depend on personal history. For example, the experience of color blue may reflect our living environment, our mood at certain time period, or other things involving individual experiences. If one grew up in the proximity to ocean, then the color blue is associated with his/her childhood memory with the ocean. If someone lives in the open space of countryside, thus blue sky is the constant companion, then the color blue may represent openness and spaciousness. On the other hand, if one is often depressed, he or she may feel sad whenever the blue color shows up[7]. I.A., instead, is independent of personal history, culture, and our stereotype, as the ability to be aware is universal among all living things.

## 6. Conclusions and the Final Comments

From these discussions we can reach the following main conclusions: The intrinsic awareness (IA) that we all have is a universal phenomenon among all living beings. I.A. is the ground state of our conscious experiences that is timeless and irreducible. The non-applicability of reductionism to study of I.A. challenges the current "theories of everything" in which consciousness is ignored.

Recently, Rowlands has developed a new theory, "universal computational rewrite system", with significant applications in particle physics and cosmology [10]. Basically, Rowlands can generate all mathematics, structure of nature, including space and time, from a zero totality. It seems that this is a unified theory that is relative simple and promising, as far as physical world is concerned. However, it is not clear how consciousness and the fundamental awareness can come out of this universal rewrite system. Although I.A. is beyond concept and space-time, consciousness, as function of I.A., can be and should be addressed by any theory that is complete and unifying. Linde has proposed that consciousness should be one of the fundamental variables, such as space, time, matter, in any unifying theory [11].

---

[7] In American idiom, "feeling blue" means feeling depressed or unhappy.

More than thirty years ago, some physicists have speculated that there are parallels between modern physics and Eastern philosophies, even suggesting that vacuum state coming out of the quantum field theory of modern physics resembles the concept of emptiness in Buddhism [12, 13]. Recently, this parallel was revisited by Buddhist scholar Wallace [14]. Even though there are some similarities between the concept of ultimate reality in Eastern thought and the basic state of nature in physics, and these similarities will be further explored, the vacuum state addressed in quantum field theory, at least in the current form, does not explicitly involve conscious mind.

I.A. is different from the "field" in the unified field model for consciousness based on the "coherent field" developed when many people practicing meditation together [15-17]. This "coherent, unified consciousness" was suggested being similar to the "field" concept in physics. First of all, the "pure consciousness state" discussed in these articles is not necessarily the intrinsic awareness. Just because one's mind is calm, quiet, and no thoughts does not guarantee it is in the intrinsic awareness as defined in this work. Secondly, the so-called "unified, global conscious field" requires many people to establish; while each one of us has I.A. with us; there is no need for anyone else's presence to experience I.A.

## 7. Acknowledgement

The author thanks Dr. Gilles Nibart for his encouragement and suggestions for references.

## References:

1. For a review on the problems encountered by the studies of consciousness, please see: "Philosophy of Mind: Classical and Contemporary Readings", Ed. by David Chalmers, Oxford University Press, 2002, and references therein.
2. Chalmers, David, "Facing Up the Problem of Consciousness", Journal of Consciousness Studies, vol. 2, pp. 200-219, 1995; David Chalmers, "The Puzzle of Conscious experience", Scientific American Dec 1995.
3. Sheng, Yen, "Illuminating Silence", Watkins Publishing, London, 2002.
4. Suzuki, S., "Zen Mind, Beginner's Mind", Weatherhill, 2005.
5. Nyima, Chokyi, "The Union of Mahamudra and Dzogchen", Rangjung, Yeshe Publications, 1986.
6. Xuan, Hua, "Shurangama Sutra", Dharma Realm Buddhist Association, 1992.
7. One of the earliest concepts of space involves the "void" between the discrete

numbers. See, for example, Aristotle, Metaphysics, 1080 b 33.
8. Boring, Edwin G., "A history of introspection", Psychological Bulletin, vol. 50 (3), pp. 169–189, 1953.
9. For a comprehensive review of research on contemplative practices, see: Shapiro, Shauna, Walsh, Roger, and Britton, Willoughby B., "An Analysis of Recent Meditation Research and Suggestions for Future Directions", J. for Meditation and Meditation Research, vol. 3, pp. 69-90, 2003; and references therein.
10. Rowlands, Peter, "From Zero to Infinity", World Scientific, 2007.
11. Linde, Andrei, "Particle Physics and Inflationary Cosmology", Harwood Academic Publishers, 1990.
12. Capra, Fritjof, "Tao of Physics", Shambhala Publications, 1976.
13. Zukav, Gary, "Dancing Wuli Master", William Morow and Company, 1979.
14. Wallace, Alan, "Vacuum State of Consciousness: A Tibetan Buddhist View" presented at the 5th Biennial International Symposium of Science, Technics and Aesthetics: Space, time, and Beyond," Lucerne, Switzerland, January 19, 2003. http://www.neugalu.ch/english.htm.
15. Orme-Johnson, DW. Dillbeck, MC. Wallace, RK., and Landrith, GS. "Intersubject EEG coherence: is consciousness a field?" International Journal of Neuroscience, vol. 16(3), pp. 203-209, 1982.
16. Hagelin, J., "Is consciousness the unified field? A field theorist's perspective", Modern Science and Vedic Science, vol. 1, pp. 29-87, 1987.
17. Pockett, Susan, "Field theories of consciousness/Field theories of global consciousness" And references therein. Scholarpedia, ISPN 1941-6016, 2009. http://www.scholarpedia.org/article/Field_theories_of_consciousness/Field_theories_of_global_consciousness.